\documentclass[a4paper,10pt]{article}

\usepackage[spacing]{microtype}
\usepackage[T1]{fontenc}
\usepackage{amssymb}
\usepackage{amsfonts}
\usepackage{textcomp}
\usepackage{amsthm,bbm}
\usepackage{textcase}
\usepackage{xcolor,hyperref}
\usepackage[round]{natbib}

\definecolor{DarkRed}{rgb}{0.4,0.1,0}
\definecolor{MidRed}{rgb}{0.6,0,0}
\hypersetup{colorlinks,citecolor=black,urlcolor=MidRed,linkcolor=MidRed}

\newcommand{\citepos}[1]{\citeauthor{#1}'s \citeyearpar{#1}\xspace}
\let\cite=\citep

\usepackage{tools}
\usepackage{listings}
\usepackage{stmaryrd}

\definecolor{ListingBar}{rgb}{0.8,0.8,0.8}
\lstnewenvironment{prolog} {\lstset{%
	language=prolog,
	columns=spaceflexible,
	identifierstyle=\itshape,
	emphstyle=\upshape,
	xleftmargin=1.2em,
	emph={X,Y,Z},
	morekeywords={table}
	}} {}
\lstnewenvironment{ocaml} {\lstset{language=[Objective]Caml,
	columns=spaceflexible,
	frame=lines,
	aboveskip=0.75em,
	belowskip=0.75em}} {}
\lstnewenvironment{ocamlet} {\lstset{language=[Objective]Caml,columns=spaceflexible,xleftmargin=1.2em}} {}
\lstnewenvironment{badocaml} {\lstset{
	frame=leftline,
	framesep=0.9em,
	framerule=0.3em,
	xleftmargin=1.2em,
	columns=spaceflexible,
	rulecolor=\color{ListingBar},
	language=[Objective]Caml
	}} {}
\lstnewenvironment{ocaml-tt} {\lstset{language=[Objective]Caml,
  basicstyle=\ttfamily\small,
	emphstyle=\mdseries,
	keywordstyle=\bfseries,
  identifierstyle=\upshape, 
	columns=spaceflexible,
	frame=single,
	aboveskip=0.75em,
	belowskip=0.75em}} {}

\def\OCaml{OCaml\xspace}
\def\TheAcknowledgments{The author would like to acknowledge the support of Nicolas Gold, UCL, and
the UK Engineering and Physical Sciences Research Council (grant number EP/G060525/2)
during the period when most of this manuscript was written in 2013, and the support of Jukedeck Ltd.
while it was being completed in 2017.}
\def\TheAuthors{Samer Abdallah (\textsf{samer.abdallah.00@gmail.com)}}
\def\TheTitle{Memoisation: Purely, Left-recursively,\\and with (Continuation Passing) Style}
\def\TheAbstract{Memoisation, or tabling, is a well-known technique that yields large improvements
in the performance of some recursive computations. Tabled resolution in Prologs such
as XSB and B-Prolog can transform so called \emph{left-recursive} predicates from 
non-terminating computations into finite and well-behaved ones. In the functional
programming literature, memoisation has usually been implemented in a way that does
not handle left-recursion, requiring supplementary mechanisms to prevent non-termination.
A notable exception is \citepos{Johnson1995} continuation passing approach in Scheme.
This, however, relies on mutation of a memo table data structure and coding in explicit
continuation passing style. We show how Johnson's approach can be implemented purely functionally
in a modern, strongly typed functional language (\OCaml), presented via
a monadic interface that hides the implementation details, yet providing a way
to return a compact represention of the memo tables at the end of the computation.
}

\date{14th July, 2017\footnote{The bulk of this note was written in 2013 while the author
was affiliated with UCL. Sections 7 and 8 were written in July 2017 while the author was affiliated with Jukedeck Ltd.}}

\title{\TheTitle}
\author{\TheAuthors}

\begin{document}
	\maketitle
	\begin{abstract}\TheAbstract\end{abstract}
	
\section{Introduction}
% In this technical note, we present a purely functional, strongly typed implementation
% of memoisation and nondeterminism capable of handling left-recursive
% computations and giving access to the resulting memo tables, making the method
% suitable for parsing left-recursive grammars and returning a compact representation
% of the resulting parse trees.

Memoisation \citep{Michie1968,Norvig1991} is a well known technique for speeding up
computations involving repeated copies of the same sub-problem by
storing the results of solving such sub-problems and then referring to these
stored results later rather than recomputing them, thus trading
space for time. As such, it is a form of dynamic programming, and is
especially effective for computing certain recursive functions, which may have
exponential time complexity when implemented directly but are reduced to
polynomial or linear complexity when implemented with memoisation. The classical
example is the function for finding the \nth{n} number in the Fibonacci 
sequence, which starts $0, 1, 1, 2, 3 \ldots$, with each successive number the
sum of the previous two. Hence, the \nth{n} number is
\def\fib#1{\operatorname{\textit{fib}} #1}
\begin{equation}
	\fib n = \begin{cases}
											n & \text{if $n\in\{0,1\}$} \\
											% 1 & \text{if $n=1$} \\
											\fib (n-2) + \fib (n-1) & \text{otherwise.}
										\end{cases}
\end{equation}
Implemented directly in a language supporting recursive functions, the time taken
to compute $\fib n$ grows exponentially with $n$, because very many computations
are repeated.
% . This is because the
% function calls itself very many times and often with the same argument. 
For example,
to compute $\fib 8$, we must compute $\fib 6$ and $\fib 7$. Assuming we compute
$\fib 6$ first, we must then compute $\fib 7 = \fib 5 + \fib 6$. But we have
already computed $\fib 6$, and therefore end up duplicating that computation.
Memoisation improves the situation dramatically by removing these duplicates.

The Fibonacci function is a deterministic computation. If we expand our scope to 
\emph{nondeterministic} computations, which can produce zero or more answers
\cite{Wadler1985}, then it becomes possible to define \emph{left recursive} computations that not
only call themselves recursively, but call themselves recursively \emph{with the
same arguments}. Implemented directly, such computations do not terminate.
Precisely this situation crops up when working with grammars
(such as grammars for natural languages) which include left-recursive production rules.
These can often be the most succinct way of describing certain grammatical
constructions. A left-recursive grammar can always be transformed into one
which is not left-recursive one which recognises the same language 
\cite{Moore2000,JohnsonRoark2000} but
doing so can force a more convoluted programming style and make it harder to
build a semantic representation of sentences while parsing. It may also be
impossible to do this for probabilistic grammars without affecting the resulting
distribution over sentences.
Hence, the ability to handle left recursion can be useful feature.

\makebox[0pt]{\lstMakeShortInline| \lstMakeShortInline[language=Prolog,emph={X,Y,Z},emphstyle=\upshape,identifierstyle=\itshape]@}

\subsection{Tabling in logic programming}
\label{s:memologic}
In the logic programming community, memoisation is often referred to as \emph{tabling},
and is a feature of several Prolog implementations, including XSB, B-Prolog and YAP.
By tabling the relevant predicates, both deterministic and
nondeterministic recursive predicates can be written that would be exponentially
slow or non-terminating without tabling. For example, in B-Prolog we can compute
the transitive closure of an @edge/2@ relation as follows:
\begin{prolog}
	% left recursive transitive closure of edge/2.
	:- table path/2.

	path(X,Z) :- path(X,Y), path(Y,Z).
	path(X,Z) :- edge(X,Z).

	edge(a,b).
	edge(b,c).
\end{prolog}
Prolog's standard depth-first search would immediately go into an infinite recursion
on the first clause of @path/2@, but with tabling, this presents no difficulties: 
entering a query @path(a,X)@ produces
the solutions @b@ and @c@ with no duplicates.

Given Prolog's close association with natural language applications, a
relationship between tabling and efficient chart parsing algorithms
was quickly noticed \cite{Warren1975,PereiraWarren1983}. Chart parsers build a compact representation
of the results of parsing various subsequences of the target sequence, which
can then be used to build higher levels of the syntax tree while avoid repeated
computations. These charts are essentially memo tables. The correspondence is even clearer if we
use Prolog's definite clause grammar (DCG) syntax to write the grammar,
yielding a succinct, executable specification. Without tabling, applying the
top level predicate to an input sequence executes a recursive descent, backtracking
search for a valid parses. As such it is vulnerable to exponential time complexity
or non-termination with left-recursive clauses. As \citet{Warren1995}
states,  when you write a DCG, you get a parser ``for free'', but not necessarily
a very good one, but with tabling, the parser you get ``for free''
is a pretty good one, essentially the same as Earley's algorithm \cite{Earley1970}.
Indeed, the process of tabled resolution in Prolog is sometimes called ``Earley deduction''
\cite{Porter1986}.

Thus, tabling in Prolog is a powerful tool, but in most cases, it requires
low-level support within the Prolog engine and cannot be modified by the programmer.
Attempts have been made to reduce the reliance on low-level support 
\cite{RameshChen1997,De-GuzmanCarroHermenegildo2008}, but in these cases,
a foreign library (\eg in C) is still required to achieve acceptable
performance.\footnote{Since the bulk of this note was written, a Prolog implementation
of tabling as a library \cite{DesouterVan-DoorenSchrijvers2015} based on delimited 
continuations \cite{SchrijversDemoenDesouter2013} has become available.}

\subsection{Memoisation in functional programming}
\label{s:memofun}
In the functional programming world, memoisation has been tackled in an
embedded way, by writing code in the language rather than by implementing
new primitives in the compiler or interpreter.

\citet{Norvig1991} showed how to define a higher-order function in LISP that,
given any function, produces a memoised version of it. Crucially it relied
on mutation effects, both to manage the memo tables using updatable
variables and to modify the symbol table mapping symbols to functions so that
any recursive calls in the memoised function are redirected to the memoised
version. The system was demonstrated in a program for parsing context free grammars,
using lists to represent non-determinism as described by \citet{Wadler1985}.

\citet{Johnson1995} extended this approach, this time working in Scheme, introducing
what are essentially parser combinators, and transforming the program into
a continuation passing style (CPS). Having done this, Johnson showed how the memoisation
process could accommodate left-recursion by effectively suspending the branches of the
computation which are left recursive and allowing the other branches to proceed until
they produce a result, at which point, suspended branches can resume and consume
the newly generated result. In this approach, non-determinism is not represented with
lists of results, but by calling a captured continuation multiple times, once with
each alternative. Notably, it essentially the same as the most commonly used tabling
strategies in Prolog, OLDT resolution \citep{TamakiSato1986} and SLG resolution \cite{ChenWarren1993}.

\citet{Frost1994}
showed how a memoising combinator can be defined in a 
pure lazy functional language (Miranda, a precursor of Haskell) by explicitly
threading a state (containing the memo tables) through all memoised computations.
The framework was applied to defining functional parsers and parser combinators,
this time using lists to represent non-determinism, but could not handle left 
recursive grammars or left-recursion in general. In subsequent publications,
Frost and his co-workers developed the idea, adopting a monadic approach to state threading,
this time in Haskell \cite{Frost2003}, developing a way to handle left-recursive
grammars, though not by using Johnson's method, but by managing and limiting the depth 
of left-recursive calls \cite{FrostHafiz2006}, and adding the ability to return
a compact representation of multiple parse trees \cite{FrostHafizCallaghan2007,FrostHafizCallaghan2008}.
Frost \etal acknowledge Johnson's continuation passing approach but do not adopt it;
indeed, they express surprise that it has not received much attention and suggest
that that this may be because the approach is ``somewhat convoluted and extending 
it to return packed representations of parse trees [\ldots] could be too complicated.''

% The only other example of memoisation in CPS that we are
% aware of is the work of \citet{SwadiTahaKiselyov2006}, but here CPS is used only
% to increase the efficiency of stateful operations, and not to implement nondeterminism;
% the authors do not address left-recursive computations.

In this note, we show how Johnson's CPS memoisation solution can indeed be
expressed in a functional style without relying on any mutable
state or side effects to any significant degree. 
We hide the implementation behind a monadic interface that provides
not only memoisation, but also nondeterminism, as computation effects.
With a slight complication of the 
interface, we can also return a compact representation of the memo tables 
built during parsing, which can then
be interrogated to produce all possible parse trees.
The code presented below is in \OCaml, which is not strictly a pure functional
language, but we restrict ourselves to a pure functional fragment. The only exception
to this is that mutable references are used to implement a universal type, but
the visibility of this impurity is confined to a very limited scope.

In the following sections, we define a framework for using monads in \OCaml (\secrf{monads}),
deduce suitable types for memoisation and discuss open recursion and fixed point
combinators (\secrf{recursion}) before presenting our code for memoisation of left recursive functions
in a continuation passing style \secrf{contmemo}. We then extend this to allow the
memo tables to be extracted at the end of the computation (\secrf{contmemotable}).
The system is compared with other memoising parser frameworks in \secrf{comparison}
before concluding in \secrf{conclusion}.
Supporting code is provided in the appendix.

\section{Monads in \OCaml}
\label{s:monads}

As we intend to present complete working code, we start by defining some utility
functions corresponding to functions of the same name in Haskell:
\begin{ocaml}
	let id x = x               (* identity function *)
	let ( ** ) f g x = f (g x)  (* function composition *)
	let cons x xs = x :: xs     (* prepend item to list *)
	let curry f x y = f (x,y)
\end{ocaml}
It is well known that a wide variety of computational effects can be structured
using monads \cite{Wadler1992}. Unlike Haskell, \OCaml lacks a standard
monad library, so we provide some module interfaces for a few monad classes:
\begin{ocaml}
	module type MONAD = sig
		type 'a m
		val return : 'a -> 'a m
		val bind : 'a m -> ('a -> 'b m) -> 'b m
	end

	module type MONADPLUS = sig
		include MONAD
		val mzero : unit -> 'a m
		val mplus : 'a m -> 'a m -> 'a m
	end

	module type MONADREF = sig
		include MONAD
		type 'a ref
		val new_ref : 'a -> 'a ref m
		val get_ref : 'a ref -> 'a m
		val put_ref : 'a ref -> 'a -> unit m
	end
\end{ocaml}
|MONADREF| is modelled on Haskell's |MonadRef| type class, and provides operators for managing
polymorphic mutable references, much like \OCaml's built in |'a ref| type.
Given any monad implementing the primitive |return| and
|bind| functions, we can provide some useful derived functions equivalent
to those defined in the Haskell monad library:
\begin{ocaml}
	module MonadOps (M : MONAD) = struct
		open M

		let (>>=) = bind
		let (>>) m1 m2 = bind m1 (fun _ -> m2)

		let liftM op m = m >>= return ** op
		let liftM2 op m n = m >>= fun x -> n >>= (return ** op x)

		let rec mapM f = function 
			| {} -> return {}
			| x :: xs -> liftM2 cons (f x) (mapM f xs)
	end
\end{ocaml}
We will also be using monad transformers \cite{LiangHudakJones1995} to combine
monadic effects, in particular, layering nondeterminism over a base monad, for
which we use a port of the standard Haskell |ListT| monad transformer, adhering to the
|MONADPLUS| interface:
\begin{ocaml}
	module ListT (M : MONAD) = struct
		type 'a m = 'a list M.m
		
		module MO = MonadOps (M)
		open MO

		let return x = M.return {x}
		let bind m f = m >>= mapM f >>= (M.return ** List.concat)
		let lift m   = M.bind m return
		let mzero () = M.return {}
		let mplus f g = liftM2 (@) f g
	end
\end{ocaml}
This represents the result of a nondeterministic computation as a list
of possible values, with |mzero ()| denoting a computation that fails and |mplus a b|
denoting a nondeterministic choice between two computations |a| and |b|.

\section{Recursion and fixed-point combinators}
\label{s:recursion}

If we were to follow Norvig's or Johnson's approaches \cite{Norvig1991,Johnson1995},
we might try to define a high order function |memo : ('a -> 'b) -> ('a -> 'b)| that
takes an ordinary (pure) function and returns a memoised version of it. There are two
problems with this. 

Firstly, since our aim is to do purely functional memoisation, using monads to
represent computational effects, and maintaining the memo table requires state
handling effects, the memoised function must have a monadic type. In addition,
if a recursive function is to be able to call a memoised version of itself, it too
must be lifted into the monad.
This might lead us to propose 
|memo : ('a -> 'b m) -> ('a -> 'b m)| as our memoising operator, where 
where |m| is the type constructor of the monad which will carry the necessary effects.
However, if we attempt to write such a function, we find that, because functions cannot
be compared for equality, there is no way to 
% associate different memo tables with different functions to be memoised:%
implement the functions |get_table_for| and |modify_table_for| below:%
\footnote{The
grey bars to the left of this and other examples below denote non-functioning or incomplete code.}
\begin{badocaml}
	type ('a,'b) table  (* table for a function of type 'a -> 'b *)
	val lookup : 'a -> ('a,'b) table -> 'b option
	val insert : 'a -> 'b -> ('a,'b) table -> ('a,'b) table

	let memo f x = 
		get_table_for f >>= fun table -> (* NOT POSSIBLE *)
		match lookup x table with
		Some y -> return y
		None -> f x >>= fun y ->
			      modify_table_for f (insert x y) > (* NOT POSSIBLE *)
			      return y 
\end{badocaml}
One reasonable solution is to add an identifier
as a parameter to |memo|, yielding something like this:
\begin{badocaml}
	type id = string

	val memo : id -> ('a -> 'b m) -> 'a -> 'b m
	let memo id f x =  
		get_table id >>= fun table ->
		match lookup x table  with
		Some y -> return y
		None -> f x >>= fun y ->
			      mod_table id (insert x y) >>
			      return y 

	let rec fib x = 
		memo "fib" (function 
								| 0 -> return 0 
								| 1 -> return 1 
								| n -> liftM2 (+) (fib (n-1)) (fib (n-2))) x
\end{badocaml}
where, for the sake of brevity, we have used strings as identifiers. Behind
the scenes, |get_table| and |mod_table| can maintain an associative mapping 
between ids and memo tables (though doing so polymorphically
in a type-safe manner presents some difficulties).
This is essentially the approach taken by \citet{Frost2003}, who restrict
themselves to memoising parsing functions of uniform type, and so do
not have to deal with the polymorphism issue.
One potential problem is that it is now the programmer's
responsibility to keep track of the ids, most importantly to avoid
duplicates. Also, in some applications, it might become a distracting burden to have
to invent an identifier for each memoised function. What is required is an
\emph{allocation} step, where a unique identifier is generated for each function
to be memoised. Allocation of resources is an effectful operation, so the preparation
of a memoised function will itself need to be a monadic operation. Rather than
exposing the programmer to the details of resource allocation, perhaps the
safest option is to hide all of this in a memoising operator of type
|memo : ('a -> 'b m) -> ('a -> 'b m) m|.
This has two advantages over Frost's approach: (a) it is no longer possible to make 
an error handling the memo table ids and (b) each memo table can be properly
initialised, which may improve the performance of later operations. However, 
it also means we can no longer use an ordinary |let rec| binding to define 
|fib : int -> int m|.

This brings us to the second problem: with a monadic memoising combinator,
there is no recursive binding construct available that will let us refer to the
result of the computation produced by |memo| in the argument to |memo|.
The solution is to adopt the \emph{open} style of recursion and
use an explicit fixed-point combinator. For example, in open-recursive
style, the monadic Fibonacci function is
\begin{badocaml}
	val fib' : (int -> int m) -> int -> int m
	let fib' f = function | 0 -> return 0 | 1 -> return 1 
	                       | n -> liftM2 (+) (f (n-1)) (f (n-2))
\end{badocaml}
A fixed-point combinator, |fix : (('a -> 'b) -> ('a -> 'b)) -> 'a -> 'b|,
is most straightforwardly written in \OCaml
using a |let rec| binding:
\begin{ocaml}
  let fix f = (let rec fp x = f fp x in fp)
\end{ocaml}
This closes the recursion and allows the (un-memoised) Fibonacci function to be
written as |fix fib' : int -> int m|.
For defining two or more \emph{mutually} recursive functions, we need to generalise the
idea of an open-recursive function to allow the first argument to be a data structure
containing the fixed points of \emph{all} the recursive functions in the set; for example,
a \emph{dyadic} fixed-point combinator is
\begin{ocaml}
  let fix2 ( (f : ('a -> 'b) * ('c -> 'd) -> 'a -> 'b),
	           (g : ('a -> 'b) * ('c -> 'd) -> 'c -> 'd) ) =
	  let rec fp x = f (fp,gp) x
	  and     gp x = g (fp,gp) x
	  in (fp,gp)
\end{ocaml}
Note the types of the open-recursive functions |f| and |g| here: higher
arity fixed-point combinators will require correspondingly elaborated
types for the first argument of each open-recursive function.
With a few tricks \cite{Kiselyov2003-fixedpoint} one can also write variadic 
fix-point combinators that work with an arbitrary number of functions, but
we will not pursue that here. 

\section{Monadic memoisation}
\label{memoisation}

Using open-recursion, it is relatively easy to write memoising fixed-point
combinators that prepare the memo tables and tie up the recursion while
inserting the appropriate code for checking the memo tables on each call. 
Instead of doing this, we will follow
\citet{McAdam1997} and consider transformations of open-recursive functions
(or ``functionals'', as McAdam calls them). In this scheme,
a ``wrapper'' is a high-order function that takes an open-recursive
function and returns a new one that may do something interesting to intervene in the
operation of the original function each time it is called. McAdam showed how
memoisation can be handled by a wrapper, leaving the job of tying up the recursion
to a separate fixed-point combinator.

We make two modifications to McAdam's idea. Firstly,
to accommodate the possibility of memoising mutually recursive functions, we
generalise the type of open-recursive functions to |'c -> 'a -> 'b|, where
|'c| is the type of the data structure holding the fixed points of all
the relevant open-recursive functions, as discussed in the previous section.
Incidentally, non-recursive functions can be accommodated by setting
|'c = unit|. 
Secondly, since we are not using mutable data structures to manage the memo tables,
the creation of a memoising wrapper needs to be a monadic operation of type
|('c -> 'a -> 'b m) -> ('c -> 'a -> 'b m) m|.
Armed with such a wrapper, any number of mutually recursive functions
can be memoised using the appropriate fixed-point combinators. A simple
memoising framework equivalent to the one we have been attempting to write
above can now be sketched out:
\vspace{0.75em}
\begin{badocaml}
	type ('a,'b) id

	val new_memo : ('a,'b) id m
	val get_table : ('a,'b) id -> ('a,'b) table m
	val mod_table : ('a,'b) id -> (('a,'b) table -> ('a,'b) table) -> unit m
	val memo : ('c -> 'a -> 'b m) -> ('c -> 'a -> 'b m) m

	let memo fn = 
		new_memo >>= fun id ->
		return (fun fp x -> 
			get_table id >>= fun table ->
			match lookup x table with
			| Some y -> return y
			| None -> fn fp x >>= fun y ->
								mod_table id (insert x y) >>
								return y) 

	let test_fib n = memo fib' >>= fun f -> fix f n
\end{badocaml}
\vspace{0.75em}
This leads us to propose the following |MONADMEMO| as a general interface for any memoising monad and
an accompanying functor for defining useful operations for any memoising monad, including |mem : ('a -> 'b m) -> ('a -> 'b m) m|
for memoising non-recursive functions and |memrec| to get the memoised%: (('a -> b m) -> 'a -> 'b m) -> ('a -> 'b m) m| to get the memoisied
fixed point of an open-recursive monadic computation:
\vspace{0.75em}
\begin{ocaml}
	module type MONADMEMO = sig
		include MONAD

		val memo : ('c -> 'a -> 'b m) -> ('c -> 'a -> 'b m) m
	end

	module MemoOps (M : MONADMEMO) = struct
		include MonadOps(M)
		open M

		let mem f = memo (fun () x -> f x) >>= fun mf -> return (mf ())
		let memrec f = liftM fix (memo f)
		let memrec2 (f,g) = liftM2 (curry fix2) (memo f) (memo g)
	end
\end{ocaml}
Note that |liftM| and |liftM2| are required to apply the ordinary functions |fix| and
|fix2| to the results of the monadic memoisation operator |memo|.

\section{Nondeterminism and left-recursion}
\label{s:contmemo}

Nondeterminism, using lists to represent multiple success \cite{Wadler1985},
can already be dealt with using the memoiser sketched out in the previous section,
simply by memoising functions of type |'c -> 'a -> 'b list m|. This
results in a memo table where each input of type |'a| is associated with
a list of results instead of just one, and is equivalent to the methods
of both \citet{Norvig1991} and \citet{Frost1994}.

% This generalises straightforwardly to any monad implementing nondeterminism
% \cite{Wadler1992}, of which the list monad is perhaps the simplest: all that
% is required in the present context is a module implementing the |MONADPLUS|
% signature given in \secrf{monads}. 
In order to deal with left-recursion, however,
we will lift both nondeterminism and memoisation into a continuation passing monad, 
modelled on Haskell's |ContT| monad transformer, and adapt \citepos{Johnson1995} method.
The key to this is to notice that the continuation monad provides
\emph{delimited} (or composable) continuations, which, as \citet{Filinski1994,Filinski1999} showed,
can be used to implement the computational effects of any monad or 
combination of monads. For our purposes, we will define a |ContT| functor in
\OCaml, parameterised by a fixed 
answer type and a base monad:
\begin{ocaml}
	module type TYPE = sig type t end

	module ContT (W : TYPE) (M : MONAD) = struct
		type 'a m = {{run: ('a -> W.t M.m) -> W.t M.m}}

		let return x = {{run = fun k -> k x}}
		let bind m f = {{run = fun k -> m.run (fun x -> (f x).run k)}}
		let shift f  = {{run = fun k -> (f (return ** k)).run id}}
		let lift m   = {{run = fun k -> M.bind m k}}
	end
\end{ocaml}
In addition, because of the fixed answer type ($W.t$) of |ContT|,
we will borrow \citepos{Filinski1999} |Dynamic| module implementing
a universal type to enable sufficient polymorphism when running computations
in the memoising monad:
\begin{ocaml}
	module Dynamic = struct 
		exception Dynamic
		type t = Dyn of (unit->unit)

		let newdyn () : ('a -> t) * (t -> 'a) = 
			let r = ref None in
			( (fun a -> Dyn (fun () -> r := Some a)),
				(fun (Dyn d) -> r := None; d (); 
					              match !r with
					              | Some a -> a
					              | None -> raise Dynamic))
	end
\end{ocaml}
This module uses an \OCaml reference as a channel to communicate polymorphically
across a monomorphic interface and is the only place where effectful 
\OCaml constructs are used. The rest of the program is completely insulated
from these effects and so the system can still be considered pure---alternative
implementations could use type coercions or delimited
continuations with full answer type polymorphism \cite{AsaiKameyama2007}.

Using |ContT|, we can write a functor |MemoT| parameterised by 
an arbitrary monad of type |MONADREF| for providing typed references. This
module provides memoisation and nondeterminism using the |ListT| monad transformer.
Generalising it to use an arbitrary monad transformer for nondeterminism would
be relatively straightforward.
\begin{ocaml}
	module MemoT (Ref : MONADREF) : sig
		type 'a m

		val memo  : ('a -> 'b -> 'c m) -> ('a -> 'b -> 'c m) m
		val return : 'a -> 'a m
		val bind   : 'a m -> ('a -> 'b m) -> 'b m
		val mzero  : unit -> 'a m
		val mplus  : 'a m -> 'a m -> 'a m
		val run    : 'a m -> 'a list Ref.m

	end = struct
		module ND = ListT (Ref)
		module CC = ContT (Dynamic) (ND)
    module CO = MonadOps (CC)
		include CC

		let run m =
			let (ind,outd) = Dynamic.newdyn () in
			ND.bind (m.run (ND.return ** ind)) (ND.return ** outd)

		let liftRef m  = lift (ND.lift m)
		let mzero ()  = {{run = fun k -> ND.mzero ()}}
		let mplus f g = {{run = fun k -> ND.mplus (f.run k) (g.run k)}}
		let msum      = List.fold_left mplus (mzero ())

		let memo fop = let open CO in
			liftRef (Ref.new_ref BatMap.empty) >>= (fun loc ->
			let update x e t = liftRef (Ref.put_ref loc (BatMap.add x e t)) in
			return (fun p x ->
				liftRef (Ref.get_ref loc) >>= fun table ->
				try let (res,conts) = BatMap.find x table in
					shift (fun k -> update x (res,k::conts) table >>
													msum (List.map k res))
				with Not_found ->
					shift (fun k -> update x ({},{k}) table >>
													fop p x >>= fun y ->
													liftRef (Ref.get_ref loc) >>= fun table' ->
													let (res,conts) = BatMap.find x table' in
													if List.mem y res then mzero ()
													else update x (y::res,conts) table' >>
															 msum (List.map (fun k -> k y) conts))))
	end
\end{ocaml}
Some comments on the code are appropriate here: the basic framework is
a stack of two monad transformers on a base monad of type |MONADREF|, which
provides mutable references for storing the memo tables. The stateful
operations from |MONADREF| are lifted through the two layers using |liftRef|. 
The module |ND| provides nondeterminism layered over state such that the state
is shared across alternative branches of execution. However, this nondeterminism
is exposed (via |mplus| and |mzero|) by capturing the current continuation, using 
it twice or not at all, and combining the results using operations from |ND|.

The memo tables are implemented using the polymorphic map module
from \OCaml With Batteries; for a function of type |'a -> 'b m|, the memo
table is of type |('a, 'b list * ('b -> Dyn.t m) list) BatMap.t|.
% |'b delcont = | is the type of the delimited continuation 
% |k| captured by the control operator |shift|. 

When |memo| is applied to open-recursive function |fop|, a new, empty memo 
table is allocated and a function
implementing the memoised computation is returned. This 
retrieves the memo table and attempts to look up the argument |x|. If it is
found, |shift| is used to capture the continuation, which is
added to the list of continuations associated
with |x| in the memo table and then called for each result in the memo table entry,
combining the results using |msum|. 

If |x| is not found in the table, meaning this is the first time the memoised 
function has been applied to |x|,
a table entry is created, containing no results and one continuation,
after which |fop| is called. Then, for each result produced by |fop| that
is not already in the memo table,
the entry is updated and all the continuations registered 
for |x| are called with the new value, with the results of each again combined using |msum|.

We can try out the module on the Fibonacci function and the transitive closure 
program given in \secrf{memologic}.
\begin{ocaml}
	module Fibonacci (M : MONAD) = struct
		include MonadOps (M)
		let fib f = function | 0 -> M.return 0 | 1 -> M.return 1
												 | n -> liftM2 (+) (f (n-1)) (f (n-2))
	end

	module TransClose (M : MONADPLUS) = struct
		open M
		let edge = function | "a" -> return "b"
												| "b" -> return "c"
												|  _  -> mzero ()
		let path p x = mplus (edge x) (bind (p x) p) 
	end

	module Test = struct
		module MM = MemoT (Ref)
		module FF = Fibonacci (MM)
		module TC = TransClose (MM)
		module MO = MemoOps (MM)
		open MO

		let test_fib n = Ref.run (MM.run (memrec FF.fib >>= fun fib -> fib n))
		let test_path x = Ref.run (MM.run (memrec TC.path>>= fun path -> path x))
	end
\end{ocaml}
The |Ref| monad given in the appendix is used to provide
mutable references.
Using the run function from |MemoT| yields a computation in the |Ref|
monad which is then run using |Ref.run|: in |test_fib|, this returns a single
integer, while in |test_path|, it returns a list of strings, \eg,
|Test.test_path "a"| returns |{"b", "c"}|.

\section{Getting access to the memo tables}
\label{s:contmemotable}

If the memoising monad |MemoT| is used for parsing, the resulting memo
tables contain all the information held in the charts used by efficient chart parsing algorithms,
providing a compact representation of all the parse trees. As it
stands, there is no way to get hold of them---they are buried inside the |Ref| monad.
One way to obtain them is to modify the |memo| combinator so
that, as well as returning the memoised function, it also returns a monadic 
operator to return the current memo table for that function. Suppose that,
for a memoised function of type |'a -> 'b m|, the type of the memo table is
|('a,'b) table|. Then we might try 
|memo : ('c -> 'a -> 'b m) -> ( ('a,'b) table m * ('a -> 'b m)) m|, where
second element of the pair produced by |memo| is the memoised function as before,
but the first is a computation that returns the memo table.
This will not do: since the |MemoT|
monad includes nondeterminism as an effect, a parsing computation which ends
with reading and returning a memo table will result in \emph{multiple} memo tables,
one for each successful parse, even though the memo table is shared across
nondeterministic alternatives. 
% Consider the following code fragment, in a hypothetical extension
% of |MemoT| where |memo| produces a table reader as well as a memoised 
% function, applied to the open-recursive |path| funtion from the previous
% example:
% \begin{badocaml}
% 	memo TC.path >>= fun (get_table,memo_pathp) -> 
% 	fix memo_path 'a' >>= fun x ->
% 	get_tables >>= fun table ->
% 	return (x,table)
% \end{badocaml}
% The type of this expression is |(string * (string,string) table) m|. If
% we run this in a nondetermistic memoising monad, then the end result
% is of type |(string * (string,string) table) list|. Thus we get a \emph{list}
% of memo tables even though the state of the memo tables is shared across
% all the nondetermistic alternative.

Instead we need memo table initialisation and extraction to
to operate in the base |Ref| monad, rather than in the nondeterministic 
|ContT| monad. The work-flow will consist of preparing
memoised functions in the |Ref| monad, running the nondeterministic 
computation in the memoising monad layered over |Ref|, and then retrieving
the memo tables after dropping back into the |Ref| monad. A suitable
interface for managing this is |MONADMEMOTABLE|, along with
an accompanying functor implementing memoising fixed point operators:
\begin{ocaml}
	module type MONADMEMOTABLE = sig
		include MONAD
		module Nondet : MONADPLUS

		type ('a,'b) table = ('a * 'b list) list
		val run : 'a Nondet.m -> 'a list m
		val memo : ('c -> 'a -> 'b Nondet.m) -> 
							 (('a,'b) table m * ('c -> 'a -> 'b Nondet.m) ) m
	end

	module MemoTabOps (M : MONADMEMOTABLE) = struct
		module MO = MonadOps (M)
		open M
		open MO

		let mem f = memo (fun () x -> f x) >>= fun (g,mf) -> return (g,mf ())
		let memrec f = memo f >>= fun (g,mf) -> return (g,fix mf)
		let memrec2 (f,g) = memo f >>= fun (get_f,mf) ->
		                     memo g >>= fun (get_g,mg) ->
		                     let (fp,gp) = fix2 (mf,mg) in
		                     return ((get_f,get_g),(fp,gp))
	end
\end{ocaml}
For the sake of concreteness, we have fixed the representations
of nondeterministic alternatives and the memo tables to use lists,
but it would also be possible to parameterise the module type over alternative
representations. The |MONADMEMOTABLE| signature presents \emph{two}
monadic interfaces: the nested |Nondet| module is for use by memoised
nondeterministic computations, while the outer one provides the
memoising operator and a function to run a memoised computation.
The implementation is similar to |MemoT| except for this repackaging
into an outer and an inner module. The inner module |Nonedet| implements
the |MONADPLUS| interface using continuations. The only substantive addition is the
operator to return a memo table: the |BatMap| used internally is
converted to a list of pairs, and the
|sanitize| function removes the list of continuations associated with each 
entry before returning it.
\begin{ocaml}
	module MemoTabT (Ref : MONADREF) = struct
		module ND = ListT (Ref)
		include Ref

		module Nondet = struct
			include ContT (Dynamic) (ND)
			let mzero ()  = {{run = fun k -> ND.mzero ()}}
			let mplus f g = {{run = fun k -> ND.mplus (f.run k) (g.run k)}}
		end

		module RefO = MonadOps (Ref)
		module CCO = MonadOps (Nondet)

		type ('a,'b) table = ('a * 'b list) list

		let run (m : 'a Nondet.m) : 'a list Ref.m =
			let (ind,outd) = Dynamic.newdyn () in
			ND.bind (m.run (ND.return ** ind)) (ND.return ** outd)

		let memo fop = let open RefO in 
			new_ref BatMap.empty >>= (fun loc ->

			let liftRef m  = Nondet.lift (ND.lift m) in
			let sanitize (x,(s,_)) = (x, BatSet.fold cons s {} in
			let update x e t = liftRef (put_ref loc (BatMap.add x e t)) in

			return (
				get_ref loc >>= return ** List.map sanitize ** BatMap.bindings, 
				( fun p x -> let open CCO in let open Nondet in
						liftRef (get_ref loc) >>= fun table ->
						try let (res,conts) = BatMap.find x table in
							shift (fun k -> update x (res,k::conts) table >>
															BatSet.fold (mplus ** k) res (mzero ()))
						with Not_found ->
							shift (fun k -> update x (BatSet.empty,{k}) table >>
															fop p x >>= fun y ->
															liftRef (get_ref loc) >>= fun table' ->
															let (res,conts) = BatMap.find x table' in
															if BatSet.mem y res then mzero ()
															else update x (BatSet.add y res,conts) table' >>
																	 List.fold_right (fun k -> mplus (k y)) 
																									 conts (mzero ())))))
	end
\end{ocaml}
Note that the results are now being collected in a data structure optimised for fast lookups (|BatSet|).
We can use this module to implement a simple parser which, thanks to
memoisation, is equivalent to an Earley parser. First, we need a parser
combinator library \cite{Hutton1990}: parsers are represented as nondeterministic
monadic computations of type |'a list -> 'a list m|, taking a list of tokens to 
parse and producing, if successful, the list of remaining tokens. The operators 
|*>| and |<|>| combine two parsers, in sequence or as alternatives, respectively.
The primitive |epsilon| parses an empty sequence and |term x| matches a single token
|x|.
\begin{ocaml}
	module Parser (M : MONADPLUS) = struct
		let ( *> ) f g xs = M.bind (f xs) g
		let ( <|> ) f g xs = M.mplus (f xs) (g xs)
		let epsilon xs = M.return xs
		let term x = function | y::ys when x=y -> M.return ys
													 | _ -> M.mzero ()
	end
\end{ocaml}
Then we can write a small grammar, equivalent to \citepos{Johnson1995} 
example of a left recursive grammar.
\vspace{0.5em}
\begin{ocaml}
	module Test2 = struct
		module MM = MemoTabT (Ref)
		include Parser (MM.Nondet)
		include MonadOps (MM)
		include MemoTabOps (MM)
		open MM

		let v   = term "likes" <|> term "knows" 
		let pn  = term "Kim" <|> term "Sandy" 
		let det = term "every" <|> term "no" 
		let n   = term "student" <|> term "professor" 
		let np' = fun np -> pn <|> det *> n <|> np *> term "'s" *> n
		let vp' = fun np (vp,s) -> v *> np <|> v *> s
		let s'  = fun np (vp,s) -> np *> vp

		let success = function | {} -> true | _ -> false
		let parse input =
			memrec np' >>= fun (get_np,np) ->
			memrec2 (vp' np,s' np) >>= fun ((get_vp,get_s),(vp,s)) ->
			run (s input) >>= fun results ->
			get_s >>= fun s_memo ->
			get_np >>= fun np_memo ->
			get_vp >>= fun vp_memo ->
			return (List.exists success results, 
						  {"s",s_memo; "np",np_memo; "vp",vp_memo})
	end
\end{ocaml}
The non-recursive nonterminals |v, pn, det| and |n| match single words, while the
open-recursive nonterminals |np', vp'| and |s'| match phrases and sentences and 
are defined as functions taking the fixed points of their recursions as arguments.
Note that |np| is left recursive, to allow for noun phrases such as ``Kim's professor''.

The function |parse| returns a computation in the |Ref| monad. It ties up the 
recursive parsers using the memoising fixed point
operators |memrec| and |memrec2|, processes a list of strings and returns |true| or |false| to 
indicate whether or not the whole input was parsed, along with a list containing the
memo tables for the three recursive rules |s|, |np| and |vp|. It can be run in the 
\OCaml top-level interpreter using
|Ref.run|, for example:
\begin{ocaml-tt}
	# Ref.run (Test2.parse {"Sandy";"'s";"professor";"knows";"Kim"});;
	- : bool * (string * (string list, string list) Test2.MM.table) list =
	(true,
	 {("s", {({"Kim"}, {});
				   ({"Sandy"; "'s"; "professor"; "knows"; "Kim"}, { {} })});
	  ("np", {({"Kim"}, { {} });
				    ({"Sandy"; "'s"; "professor"; "knows"; "Kim"},
				     { {"knows"; "Kim"}; 
				       {"'s"; "professor"; "knows"; "Kim"} })});
	  ("vp", {({}, {}); 
				    ({"'s"; "professor"; "knows"; "Kim"}, {});
				    ({"knows"; "Kim"}, { {} })})})
\end{ocaml-tt}
In this case, the sentence was successfully parsed, and the memo tables show that,
for example, |{"Kim"}| could not be parsed as a sentence, but was 
parsed as a noun phrase, and also that |{"Sandy"; "'s"; "professor"; "knows"; "Kim"}|
admits of two partial parses as a noun phrase, the first consuming only the first token
|"Sandy"| and the second consuming the three tokens |{"Sandy"; "'s"; "professor"}|
and making use of the left-recursive production rule for noun phrases.

\section{Comparison with previous work}
\label{s:comparison}

Functional approaches to parsing, including parser combinators,
have been studied
for several decades \cite{Burge1975,Fairbairn1987,FrostLaunchbury1989,Hutton1990}. Both \citet{Norvig1991}
and \citet{Leermakers1993} use memoisation to improve efficiency, but Norvig
forbids left recursive rules, while Leermakers avoids the problem of left recursion
by using a `recursive ascent' strategy, sacrificing the modularity of top-down approaches
\cite{Koskimies1990}. \citepos{Johnson1995} continuation-based system,
the basis for the one developed here,
was written in Scheme without the benefit of a strong type system, and relied
on mutation side effects to manage the memo tables. It also required 
the code to be written in explicit continuation passing style, as opposed 
to using the monadic interface |ContT| described here. The possibility of a monadic interface to
parser combinators was recognised by \citet{Wadler1990}.

\citet{Lickman1995} takes a pure functional and monadic approach to parsing left recursive
grammars, including memoisation. He relies on defining a fixed point operator for recursive
parsers which is in turn defined in terms of a fixed point operator for set-to-set functions. While
mathematically elegant, the resulting implementation suffers from potentially exponential time 
complexity. In comparison, the continuation-based approach
described here computes the fixed point incrementally, since each \emph{new} 
solution from a memoised parser is fed back into its context, which may be itself if the parser is
recursive, until no more new solutions are produced.

\citet{Frost1993} proposes `guarded attribute grammars' as a way to handle
left-recursion: each left recursive rule is `guarded' by a non-left recursive recogniser,
which delimits the segment to which the left recursive parser can then safely be applied.
However, the time complexity is still exponential in the depth of the 
left recursion.
The later work of \citet{FrostHafiz2006} (see \secrf{memofun}) improves on this, reaching $O(n^4)$
time complexity in the length of the input for left recursive grammars. In comparison, the system 
described here handles left recursion without having to look ahead to the end of the input sequence 
to limit the depth of left recursion, and achieves the same $O(n^3)$ theoretical time complexity 
as Earley's chart parser. 
Another difference is that Frost et al's system
requires each memoised parser to be given a label, whereas the proposed system does not. Finally,
Frost's system is implemented in Haskell, which supports
arbitrary recursive binding constructs without any special effort, whereas in \OCaml, it was
necessary to use open recursion and explicit fixed point operators.

\citet{FrostHafizCallaghan2007} presented
some timings of their system on a small set of abstract, highly ambiguous grammars,
some involving left recursion.
The three memoised parsers, encoded below, are |sm| (recursive); |sml| (left recursive) and |smml| (composed of
two mutually recursive rules, one of which is also left recursive).
\begin{ocaml}
	module Ambig (MM: MONADMEMOTABLE) = struct
		include Parser (MM.Nondet)
		include MonadOps (MM)
		include MemoTabOps (MM)
		open MM

		let rec sentence n = function 0 -> {} | n -> "a"::sentence (n-1)

		let sm = memrec (fun sm -> term "a" *> sm *> sm <|> epsilon)
		let sml = memrec (fun sml -> sml *> sml *> term "a" <|> epsilon)
		let smml = memrec2 ((fun (smml, aux) -> smml *> aux <|> epsilon),
													 (fun (smml, aux) -> smml *> term "a"))
	end
\end{ocaml}
In the above module, each grammar is represented as a monadic operation that will
produce a memo table extractor and the memoised parser itself. All three recognise arbitrary length sequences of the
token |"a"|, which can be generated using the |sentence| function.

\begin{table}
	\begin{center}
		\begin{tabular}{l|lll|lll}
			input & \multicolumn{3}{c|}{Proposed system} & \multicolumn{3}{c}{Frost et al, 2007} \\
			length & \textit{sm} & \textit{sml} & \textit{smml} &  \textit{sm} & \textit{sml} & \textit{smml} \\[0.1em]
			\hline 
			12 & 0.001 & \textbf{0.002} & \textbf{0.002} &  \textbf{0.001} & 0.004 & 0.003 \\
			24 & 0.008 & \textbf{0.008} & \textbf{0.01}    &  \textbf{0.005} & 0.02 & 0.02 \\
			48 & 0.08 & \textbf{0.09} & \textbf{0.11}    &  \textbf{0.02} & 0.3 & 0.3 \\
			72 & 0.39 & \textbf{0.42} & \textbf{0.52}    &  \textbf{0.10} & 2.4 & 2.5 \\
			96 & 1.2 & \textbf{1.3} & \textbf{1.6}       &  \textbf{0.26} & 8.1 & 8.7 \\
		\end{tabular}
	\end{center}
	\caption{Execution times (in seconds) comparing the system presented here with that of \citet{FrostHafizCallaghan2007}
parsing sequences of the token ``a'' of various lengths (first column) using four highly ambiguous grammars. The
best performance for each test case is indicated in boldface.}
	\label{tab:timings}
\end{table}

Frost et al's Haskell code\footnote{Available at \url{http://hafiz.myweb.cs.uwindsor.ca/xsaiga/imp.html}.}
was modified to (a) reduce the effect of laziness by traversing the entire resulting data
structure (derived from the memo tables) and (b) compute the total execution time
including the final traversal but not any printing or writing to files.
The programs for both systems were fully compiled, rather than being run in the
\OCaml or Haskell interactive environments. The results, obtained using
a 2012 Macbook Pro with a 2.5 GHz Intel Core i5 CPU and 8 GB of memory, are shown
in Table~\ref{tab:timings}. The overall picture that emerges is that Frost et al's system
performs very well for non left recursive grammars, partly, one suspects, due to Haskell's laziness and
sophisticated optimising compiler, which can eliminate many the overheads associated with
data structures and high-order function manipulations, but there may be other factors,
such as the use of integers to represent the parsing state (\ie, the index of the next token to
be processed) as opposed to using the tail of the input sequence in the present system.
However, in all cases, the continuation based system handles 
left recursion more effectively.

% This implementation: 190 lines of code, including the \OCaml implementations
% of standard Haskell monads.
% 98 without haskell standard stuff

% Parser: 278 lines, inc 15 for state monad
% minimal: 164 - 11 for state
\section{Conclusions}
\label{s:conclusion}

A purely functional, continuation-based system for memoising recursive and left 
recursive nondeterministic computations, including those involved in parsing
left recursive grammars, has been presented in the form of a complete implementation
in the functional programming language \OCaml. The three computational effects 
required: statefulness, nondeterminism and delimited continuation capture, were
implemented as a stack of monads,
The system was compared with that of \citet{FrostHafizCallaghan2007},
which shares many of the same aims and tools, but uses a different method to handle
left recursion, as well as differing in a number of other respects, such as the need to
assign labels to memoised parsers and the representation of the parser state.
It was found that the continuation based system was more efficient asymptotically
for left recursive grammars, but was slower for shorter input sequences and non-left
recursive grammars, possibly due to the overheads introduced by the stack of monads.
Hence, one strategy for improving the performance is to eliminate the stack of monads and investigate the
use of delimited continuations as a primitive mechanism and implement the 
computational effects directly in delimited control operators.

\citet{Filinski1994} showed the close relationship between monads and delimited 
continuations: the present system relies on a monad to represent
delimited control operators, as previously by \citet{DyvbigJonesSabry2005}. 
Conversely, it is also possible to implement layered monadic effects using delimited control
as a primitive \cite{Filinski1999}. \citet{Kiselyov2012} describes his |delimcc| library
(written in 2001) which implements delimited continuations for \OCaml efficiently
in so-called `direct style', that is, without introducing the data structures
and associated overhead required in the monadic approach. The system described
here could easily be transformed into a `direct style' implementation, using
delimited control operators to implement the required effects (statefulness,
nondeterminism and memoisation) when required, but executing ordinary \OCaml
code directly and without overhead for the pure functional parts of the computation.

All the code used to generate the results in this note, including the modified
version of Frost et al's code, is available at \url{http:github.com/samer--/cpsmemo}.

\vspace{1em}
\bigskip
\noindent
\textbf{Acknowledgments}\\
\TheAcknowledgments

\appendix
\section{Implementation of supporting modules}
To implement |MONADREF| without using \OCaml mutable references,
we use a state monad where the state is a polymorphic key-value store implemented
using the |BatMap| functor from \OCaml With Batteries, with integer-valued keys.
This uses a balanced binary tree internally, with
lookup and insertion costs of $O(\log n)$ where $n$ is the number of items
in the store. The key type |'a loc| is opaque to users of the module.
The following code fragment must be included between the definitions of the
|ContT| and |MemoT| functors.
\begin{ocaml}
	module Store : sig
		type t
		type 'a loc

		val empty : t
		val new_loc : t -> 'a loc * t
		val get : 'a loc -> t -> 'a 
		val put : 'a loc -> 'a -> t -> t
		val upd : 'a loc -> ('a -> 'a) -> t -> t

	end = struct
		module M = BatMap.Make(BatInt)

		type t =  int * Dynamic.t M.t
		type 'a loc = int * ('a -> Dynamic.t) * (Dynamic.t -> 'a)

		let empty = (0,M.empty)
		let get (j,_,outd) (_,m)     = outd (M.find j m) 
		let put (j,ind,_) x (i,m)    = (i,M.add j (ind x) m)
		let upd (j,ind,outd) f (i,m) = (i,M.modify j (ind ** f ** outd) m)
		let new_loc (i,m) = let (ind,outd) = Dynamic.newdyn () in 
                        ((i,ind,outd),(i+1,m))
	end

	module StateM (State : TYPE) = struct
		type 'a m = State.t -> 'a * State.t 
		type state = State.t

		let return x s = (x,s)
		let bind m f s = let (x,s')=m s in f x s'
		let get s   = (s,s)
		let put s _ = ((),s)
		let upd f s = ((),f s)
	end

	module Ref = struct
		include StateM(Store)
		type 'a ref = 'a Store.loc

		let put_ref loc x  = upd (Store.put loc x)
		let upd_ref loc f  = upd (Store.upd loc f)
		let get_ref loc    = bind get (return ** Store.get loc)
		let new_ref x = bind Store.new_loc (fun loc ->
										bind (put_ref loc x) (fun _ -> return loc))
		let run m = fst (m Store.empty)
	end
\end{ocaml}
% vim: ts=2 sw=2

	%\pagebreak
	\bibliographystyle{abbrvnat}
	{\bibliography{all,c4dm,compsci}}
\end{document}